\documentclass[conference]{IEEEtran}
\usepackage[hyphens]{url}
\usepackage{hyperref}
\hypersetup{colorlinks,allcolors=blue}
\usepackage{doi}
\usepackage{xurl}

\usepackage[backend=bibtex,style=ieee,citestyle=numeric-comp]{biblatex}
\usepackage{enumitem}
\usepackage{array}
\usepackage{makecell}
\usepackage{graphicx}
\graphicspath{{images/}}
\usepackage{comment}
\usepackage{amsmath}
\usepackage[normalem]{ulem}
\usepackage[T1]{fontenc}
 
\usepackage{xcolor} 
\usepackage{balance}

\begin{document}
\title{Weak Enforcement and Low Compliance in\\
  PCI DSS: A Comparative Security Study}

\author{
    \IEEEauthorblockN{Soonwon Park and John D. Hastings}
    \IEEEauthorblockA{The Beacom College of Computer and Cyber Sciences, Dakota State University\\ soonwon.park@trojans.dsu.edu, john.hastings@dsu.edu
    }
}

\maketitle

\begin{abstract}
Although credit and debit card data continue to be a prime target for attackers, organizational adherence to the Payment Card Industry Data Security Standard (PCI DSS) remains surprisingly low. Despite prior work showing that PCI DSS can reduce card fraud, only 32.4\% of organizations were fully compliant in 2022, suggesting possible deficiencies in enforcement mechanisms. This study employs a comparative analysis (qualitative and indicator-based) to examine how enforcement mechanisms relate to implementation success in PCI DSS in relation to HIPAA, NIS2, and GDPR. The analysis reveals that PCI DSS significantly lags far behind these security frameworks and that its sanctions are orders of magnitude smaller than those under GDPR and NIS2. The findings indicate a positive association between stronger, multi-modal enforcement (including public disclosure, license actions, and imprisonment) and higher implementation rates, and highlight the structural weakness of PCI DSS's bank-dependent monitoring model. Enhanced non-monetary sanctions and the creation of an independent supervisory authority are recommended to increase transparency, reduce conflicts of interest, and improve PCI DSS compliance without discouraging card acceptance.
\end{abstract}


\section{Introduction}

The ease of use of credit and debit cards and their widespread acceptance by organizations (merchants) have led to the vast majority of payments today being processed by means other than cash. According to the Federal Reserve Bank of San Francisco, 57\% of retail transactions are done with credit or debit cards \cite{federal_2022findings_2022}. However, the security of those transactions has been an ongoing concern. 63\% of credit and debit card holders have been fraud victims, or about 134 million Americans \cite{securityorg_2025}.

The Payment Card Industry Data Security Standard (PCI DSS) is designed to encourage businesses to protect payment card details, and this improves security against data breaches. Studies have shown the effectiveness of PCI DSS in reducing credit card fraud \cite{siraj_evaluating_2023,nicho_anintegrated_2013}. However, despite the clear benefits, only 32.4\% of organizations are fully PCI DSS compliant \cite{verizon_2024payment}.

This raises a key question: How can compliance with PCI DSS be improved? Specifically, it is important to consider whether this low compliance rate is unique to PCI DSS, or if similar patterns are observed in other similar frameworks. If other data security frameworks such as HIPAA (Health Insurance Portability and Accountability Act), the Directive (EU) 2022/2555 (NIS2) and GDPR exhibit higher implementation rates, it becomes necessary to examine the factors contributing to this disparity. One plausible explanation lies in the differences in enforcement mechanisms. 

This study compares success rates of PCI DSS and other data security frameworks, namely HIPAA, NIS2, and GDPR, using implementation rates as the primary metric, with additional indicators employed when direct implementation data is unavailable. It further investigates the enforcement mechanisms in each data security framework to determine whether stronger enforcement is associated with higher compliance. The aim is to assess whether enforcement strategies influence implementation outcomes and to identify practical recommendations for improving compliance in PCI DSS. Data for the study were drawn from cited publicly available official sources and studies from 2022--2025. This investigation is guided by the following primary research question:

\begin{enumerate}[label={\textbf{RQ:}},left=1.0em]
\item Do weak enforcement mechanisms in PCI DSS cause a low success rate?
\end{enumerate}

\noindent To support exploration of this main question, 
three sub-questions are examined using a comparative analysis:
\begin{enumerate}[label={\textbf{RQ\arabic*:}},left=1.0em]
\item What enforcement mechanisms do these frameworks use?
\item How successful is PCI DSS 
compared to other frameworks?
\item What is the relationship between enforcement mechanisms and success rates?
\end{enumerate}

The paper is organized as follows. Section \ref{rq1} compares enforcement mechanisms and responsible authorities across the four frameworks (RQ1). Section \ref{rq2} evaluates the relative success rates of each framework (RQ2). Section \ref{rq3} analyzes the relationship between enforcement mechanisms and success rates, and provides recommendations for strengthening PCI DSS (RQ3). Section \ref{conclusion} concludes.

\section{Enforcement mechanism and authority comparison (RQ1)}
\label{rq1}
To address RQ1, this section compares enforcement mechanisms by examining the authorities responsible, their key characteristics, and the differences in enforcement approaches across several comparable data security frameworks. The comparable data security frameworks are HIPAA, NIS2, and GDPR. HIPAA was included to provide a U.S focused comparison: while PCI DSS protects cardholder data internationally including within the U.S., HIPAA protects health data for U.S. citizens. NIS2 and GDPR were selected due to their supranational scope, which parallels that of PCI DSS. 

\subsection{PCI DSS}
    
    \subsubsection{Authority -- PCI SSC, Acquiring Bank}
    The Payment Card Industry Security Standards Council (PCI SSC) is a global forum created to develop, disseminate, and clarify standards for payment account security by Mastercard, JCB International, Discover Financial Services, and American Express in 2006. It establishes general security requirements, approves vendors, and evaluates and certifies payment technologies.

    Organizations (the merchants) that wish to accept payment cards establish a merchant relationship with an acquiring bank\footnote{Also referred to as `acquirer', or colloquially as `merchant bank', though the latter term can be ambiguous as in the UK it commonly refers to investment banks.}. The acquiring bank and the payment card brands enter into a contractual relationship within the payment card industry chain, which allows the bank to process card transactions on behalf of the payment card brands it partners with. When a merchant enters into a contractual relationship with an acquiring bank and wants to accept payment cards, it must undergo the acquiring bank’s merchant vetting process \cite{office_merchant_2014}.

    PCI SSC does not have legal authority; however, any organization performing credit or debit card transactions is expected to comply with the PCI DSS standard     \cite{hancock_pci_2024}.

    \subsubsection{Penalty -- \$5,000-\$100,000 per month}
    PCI SSC does not publicly disclose official penalty schedules. Nevertheless, industry sources report that non-compliance penalties can range from \$5,000 to \$100,000 per month \cite{soroush_pci_2025,salihagic_counting_2024}.

\subsection{HIPAA}
    
    \subsubsection{Authority — HHS, OCR and DOJ}

        HIPAA is currently regulated by the Department of Health and Human Services (HHS) and enforced by the Office for Civil Rights (OCR) under 45 C.F.R. Parts 160 and 164, Subparts A, C, and E. OCR is responsible for investigating complaints, conducting compliance reviews, and providing education and outreach to promote adherence to the Privacy and Security Rules.
        If a complaint indicates a possible criminal violation of HIPAA, OCR may refer the case to the Department of Justice for further investigation (DOJ).
        When a covered entity fails to resolve a violation, OCR may impose civil monetary penalties \cite{usdepartment_howocr_nd}.

    \subsubsection{Penalty – Cap: \$2.1M/Year per Category (Was \$1.5M in 2018)}

        Civil penalties range from \$141 to \$71,162 per violation, with a maximum of \$2.1 million per calendar year per violation category. This annual cap applies separately to each type of HIPAA requirement violated, allowing penalties to accumulate across multiple categories \cite{usdepartment_annual_2024}. Certain violations of the Privacy Rule may also lead to criminal prosecution and imprisonment for the responsible individual. In addition, enforcement actions may be publicly disclosed by HHS \cite{usdepartment_summary_2025}, and violations may result in professional license suspension under applicable state laws \cite{federation_about_2025}.

        The largest federal HIPAA settlement to date occurred in 2018 when Anthem Inc. paid \$16 million to OCR. This federal penalty was part of \$48.2 million in total penalties for the same 2015 data breach, which included additional state-law violations \cite{usdepartment_anthem_2018}. The federal penalty exceeded the \$1.5 million per-category annual cap that was in effect at the time \cite{usdepartment_civil_2013} because OCR found violations across multiple HIPAA categories (at least four). Anthem's 2018 annual revenue was approximately \$91 billion \cite{anthem_annual_2018}.

\subsection{NIS2 Directive}

    \subsubsection{Authority -- ENISA and EU member states}

In 2016, when the European Parliament wanted to assign a supranational organization to collect reports of data breaches and computer incidents on the General Data Protection Regulation (GDPR) and the Network and Information Security (NIS) Directive, it therefore chose ENISA \cite{europeanunion_regulation_2013}.

Following the 2016 NIS1 (Directive (EU) 2016/1148), EU member states transposed NIS1 into national law, establishing the foundational cybersecurity framework that NIS2 would later build upon and replace in October 2024.

    \subsubsection{Penalty -- €10 million or 2\% of global turnover}
The NIS2 Directive introduced revised penalties for non-compliance by organizations. If a security incident occurs or an organization refuses to cooperate with authorities, NIS2 allows states to seek an injunction. As a result, organizations can be compelled to follow the state’s orders and may face fines as follows \cite{europeanunion_directive_2022}:

    \begin{itemize}[leftmargin=*]
        \item Essential entities face up to €10 million or 2\% of global turnover.
        \item Important entities face up to €7 million or 1.4\% of global turnover.
    \end{itemize}

\subsection{GDPR}
    \subsubsection{Authority -- CJEU and DPAs}
    The GDPR was proposed by the European Commission, adopted by the European Parliament and the Council of the European Union, and requires member states to establish supervisory authorities known as Data Protection Authorities (DPAs). DPAs are independent public authorities that monitor and supervise the application of the data protection law, with investigative and corrective powers  \cite{chassang_impact_2017}. 

    \subsubsection{Penalty -- €20 million or 4\%}

    A violation occurs when a security breach leads to the ``accidental or unlawful destruction, loss, alteration, unauthorized disclosure of, or access to, personal data transmitted, stored or otherwise processed'' \cite{europeandata_guidelines_2023}. Fines vary with severity  \cite{europeanpaliament_regulation_2016}:

\begin{itemize}[leftmargin=*]

\item Severe violation: Fines can reach €20 million or, for an undertaking (i.e., relevant economic entity or group), 4\% of its global turnover in the previous financial year, whichever is higher.

\item Minor violation: Fines can reach €10 million or, for an undertaking, up to 2\% of its global turnover in the previous financial year, whichever is higher.
\end{itemize}

    \noindent The largest penalty in GDPR history is the Meta case, with a fine of €1.2 billion (~\$1.3 billion USD) \cite{irish_data_2023}. In comparison, Meta’s revenue in 2023 was \$134.902 billion \cite{meta_annual_2024}.

\section{Success rate evaluation for each framework (RQ2)}
\label{rq2}

    To address RQ2, in this section, we evaluate the success rate of the PCI DSS framework compared to three alternative frameworks.
    The success rates for each framework are primarily calculated using the implementation rates. When the implementation rates are unavailable, an alternative metric is used: the rate of penalty cases to the total number of expected organizations, adjusted by a weighted value.\\
    \textit{\textbf{Penalty and Success Rates calculation:}}

    \begin{itemize}[leftmargin=*]
        \item Penalty rate: Penalty incidents $\div$ Total number of organizations.
        \item Success rate: Implementation rate, if available. Otherwise calculated by weighted penalty value: 1 – (penalty rate × 10,000)\footnote{To enhance interpretability of rare penalty events, we scale the raw rate by 10,000, expressing the result as penalties per 10,000 organizations. This practice is commonly used in public health and risk assessment to make small event frequencies more meaningful and comparable across large populations \cite{usdepartmentcdc_lession_nd}.}
    \end{itemize}

Table \ref{table:Success Rates Summary} summarizes the success rates calculations for each of the frameworks as detailed in the following subsections.

\vspace{-1.0em}
\begin{table}[h]
\centering

\caption{Success Rates Summary}
\label{table:Success Rates Summary} 
\vspace{-0.5em}
    \scriptsize
    \begin{tabular}{ m{3em}  m{1.5cm}  m{3cm}  m{1.5cm} }
    
    \hline
    &
    \makecell[c]{Implementation\\ Rates} &
    \makecell[c]{Alternative\\ Implementation Rates} & 
    \makecell[c]{Success\\ Rates} \\

    \hline
    
    PCI DSS&
    \makecell[c]{32\% \cite{verizon_2024payment}} &
    \makecell[c]{-} &
    \makecell[c]{32\%}\\
    \hline
    
    HIPAA &
    \makecell[c]{-} &
    \makecell[c]{92\%}
        &
    \makecell[c]{92\%} \\
    \hline
    
    NIS2&
    \makecell[c]{70\% \cite{noerr_nis2_2025}} &
    \makecell[c]{-} &
    \makecell[c]{70\%} \\
    \hline
    
    GDPR &
    \makecell[c]{-} &
    \makecell[c]{87\%}
    &
    \makecell[c]{87\%} \\
    \hline
    
    \end{tabular}
\end{table}
    
    \subsection{PCI DSS \texorpdfstring{$\approx$ 32\%}{≈ 32\%}}
    
According to the Verizon Payment Security Report, the implementation rate for PCI DSS has been declining overall since 2020, dropping from 43.4\% to 14.3\% in 2023 \cite{verizon_2024payment}. However, the sharp decline in 2023 is considered temporary and attributed to organizations preparing for the release of PCI DSS version 4.0. Therefore, we use the 2022 implementation rate of approximately 32\% (reported as 32.4\%) as the most recent representative data point for PCI DSS compliance.

    \subsection{HIPAA \texorpdfstring{$\approx$ 92\%}{≈ 92\%}}

    HIPAA implementation data (success rate) was calculated based on penalty rate relative to the estimated total number of organizations, resulting in a 2022 success rate around 91\% to 93\%. For the purposes of this report, we use the midpoint value of $\approx$92\% as a representative estimate. The success rate was determined using the following penalty rate calculation\footnote{The data used for the number of HIPAA penalty cases and organizations can be found in Appendix A, Tables \ref{tab:hipaa_cases} and \ref{tab:hipaa-scope-note}, respectively.}:

        \begin{itemize}[leftmargin=*]
            \item 2022: 91--93\% (91.2--92.7\%) \\
            \textit{Penalty rate: 0.0000073--0.0000088} \\ $\approx$ 22 penalty cases in 2022 $\div$ (2,500,000\text{ to }3,000,000) estimated number of organizations compliant with HIPAA \\
            \textit{Implementation rate: 0.912--0.927}\\ 
            $=1 - \big((0.0000073 \text{ to } 0.0000088)\times 10{,}000\big)$
            \item 2023: 95--96\% (94.8--95.7\%)  \\
            \textit{Penalty rate: 0.0000043--0.0000052} \\ $\approx$ 13 penalty cases in 2023 $\div$ (2,500,000\text{ to }3,000,000) estimated number of organizations compliant with HIPAA  \\
            \textit{Implementation rate: 0.948--0.957}\\ $=1 - ((0.0000043 \text{ to } 0.0000052) \times 10,000)$
        \end{itemize}

    \subsection{NIS2 \texorpdfstring{$\approx$ 70\%}{≈ 70\%}}    
    
        All EU member states were required to transpose NIS2 into national law by October 2024 \cite{europeanunion_directive_2022}. NIS2 shows 15.0\% initial implementation (4 out of 27 states \cite{europeancomission_commission_2024}) rising to 30\% by May 2025 (8 out of 27 states \cite{europeancomission_commission_2025}), with projections of around 70\% (19 out of 27 states) by end of 2025 \cite{noerr_nis2_2025}. For framework comparison purposes, the projected end-of-2025 implementation rate of around 70\% represents NIS2's expected implementation maturity.

    \subsection{GDPR \texorpdfstring{$\approx$ 87\%}{≈ 87\%}}

        According to research by Talend, only 30\% of organizations in 2018 and 42\% in 2019 were able to respond to data subject access requests within the GDPR-mandated one-month timeframe \cite{talend_themajority_2018,talend_gdpr_2019}. This low compliance rate contrasts sharply with later executive attitudes: between 90–92\% of business leaders in 2023–2024 studies expressed the belief that companies should respect online privacy \cite{piwikpro_five_2023,piwikpro_harmonizing_2024}. Based on regulatory penalty data relative to the total number of organizations, the GDPR implementation rate is estimated to have reached 87\% by 2022. The temporal progression suggests a maturation effect: the 2018 study occurred just 2–4 months after GDPR's implementation, the 2019 study at 12–18 months post-implementation, while the 2023–2024 studies reflect organizational behaviors 5–6 years after the regulation's introduction. Given the absence of comprehensive implementation data, this analysis uses the penalty-derived 87\% implementation rate from 2022 as the baseline benchmark.\\
        Success rate calculation\footnote{The actual implementation rate is expected to be higher than this figure suggests, and the penalty rate is projected to be lower, because the number of organizations required to comply with GDPR extends beyond those based in the EU. GDPR applies to any organization, regardless of location, that handles the personal data of European residents. The data used for the number of GDPR penalty cases and organizations can be found in Appendix A, Tables \ref{tab:gdpr_penalty} and \ref{tab:eu_companies}, respectively. }:

       \begin{itemize}[leftmargin=*]
            \item 2021: 87\% (86.9\%) \\
            \textit{Penalty rate}: 0.0000131 $\approx$ 463 penalty cases in 2021 $\div$ 35,430,241 organizations in 27 EU member states \\
            \textit{Implementation rate}: $0.869 = 1 - (0.0000131 \times 10,000)$
            \item 2022: 87\% (86.6\%) \\
            \textit{Penalty rate}:  0.0000134 $\approx$ 493 penalty cases in 2022 $\div$ 36,715,077 organizations in 27 EU member states \\
            \textit{Implementation rate}: $0.866 = 1 - (0.0000134 \times 10,000)$
        \end{itemize}

\section{Analysis of Enforcement Methods and Success Rates (RQ3)}
\label{rq3}

This section analyzes the relationship between enforcement mechanisms and the success rates (RQ3), with a focus on the impact of monetary penalties. We begin by assessing the business impact of financial penalties in major violation cases. Next, we explore the hypothetical maximum penalties for each framework, using a high-revenue organization (Meta) as a benchmark. Finally, we compare these penalties with the success rates evaluated in Section II and propose potential improvements for PCI DSS.

The business impact measures the financial burden of penalties relative to organizational size, calculated as the penalty amount divided by annual revenue.\\
Business impact calculation (\%):
    \begin{itemize}[leftmargin=1.5em, label={}]
       \item (Penalty $\div$ Annual Revenue) $\times$ 100
    \end{itemize}

\subsection{Potential Maximum Penalty and Business Impact}
Even when the same monetary penalty is applied, the financial impact can vary significantly depending on the size of the organization. To standardize our comparison, we assess the potential maximum penalty for a large enterprise, using Meta (annual revenue: \$134.9 billion) as a reference point. Table \ref{table:Hypothetical Maximum Penalties per Framework for an Organization} summarizes the monetary penalties under each framework.
   
\vspace{-1.0em}
\begin{table}[ht]
\centering

\caption{Hypothetical Maximum Penalties per Framework for an Organization with \$134.9B Revenue}
\label{table:Hypothetical Maximum Penalties per Framework for an Organization}
 
\vspace{-0.5em}
    \begin{tabular}{ m{0.5cm} >{\scriptsize}m{2cm}  m{0.6cm}  m{1.2cm}  >{\scriptsize}m{0.5cm} m{1.6cm} }
    
    \hline
    &    
    \makecell[c]{Monetary\\ Enforcement} &
    \makecell[c]{Business\\ Impact} &
    \makecell[c]{Potential\\ Max\\ Penalty} & 
    \makecell[c]{Potential\\ Business\\ Revenue} &
    \makecell[c]{Potential\\ Business\\ Impact} \\
    \\

    \hline
    PCI DSS&
        \makecell[c]{
        Max \$1.2M/year\\        
        (\$0.1M/month \cite{soroush_pci_2025})\\ 
    }&
    \makecell[c]{-}&
    \makecell[c]{ \textbf{\$1.2M}\\ {\scriptsize (Max penalty)}}&
    \makecell[c]{\$134.9B}&
    \makecell[c]{\textbf{0.00089\%}\\ {\scriptsize (1.2M \raisebox{0.2ex}{$\scriptstyle \div$} 134.9B)}}\\

    \hline
    HIPAA &
    \makecell[c]{
       Max \$1.5M/year\\
       per violation type \cite{usdepartment_civil_2013}
    }&
    \makecell[c]{0.018\%}&
    \makecell[c]{ \textbf{\$8.4M}\\ {\scriptsize (4 categories}\\ \raisebox{0.2ex}{$\scriptstyle \times$} {\scriptsize \$2.1M)}}
    &
    \makecell[c]{\$134.9B}&
    \makecell[c]{0.0062\%\\ {\scriptsize(8.4M \raisebox{0.2ex}{$\scriptstyle \div$} 134.9B)}}\\

    \hline
    NIS2&
    \makecell[c]{
        Max $\ge$€10M or \\
        2\% global turnover \cite{europeanunion_directive_2022}
    } &
    \makecell[c]{-}&
    \makecell[c]{ \textbf{\$2.7B}\\ {\scriptsize(2\% of 134.9B)}}&
    \makecell[c]{\$134.9B}&
    \makecell[c]{2\%\\ {\scriptsize(Max Penalty)}}\\
    
    \hline
    GDPR &
    \makecell[c]{
        Max €20M or 4\% \\
        of global turnover
    }&
    \makecell[c]{0.96\%}&
    \makecell[c]{ \textbf{\$5.4B}\\ {\scriptsize(4\% of 134.9B)}}&
    \makecell[c]{\$134.9B}&
    \makecell[c]{4\%\\ {\scriptsize(Max Penalty)}}\\

    \hline
    \end{tabular}
\end{table}

\subsection{Enforcement Methods, Maximum Penalties, and Success Rates Comparison (RQ3)}

\begin{figure}[ht]
        \centering
        \includegraphics[width=0.48\textwidth]{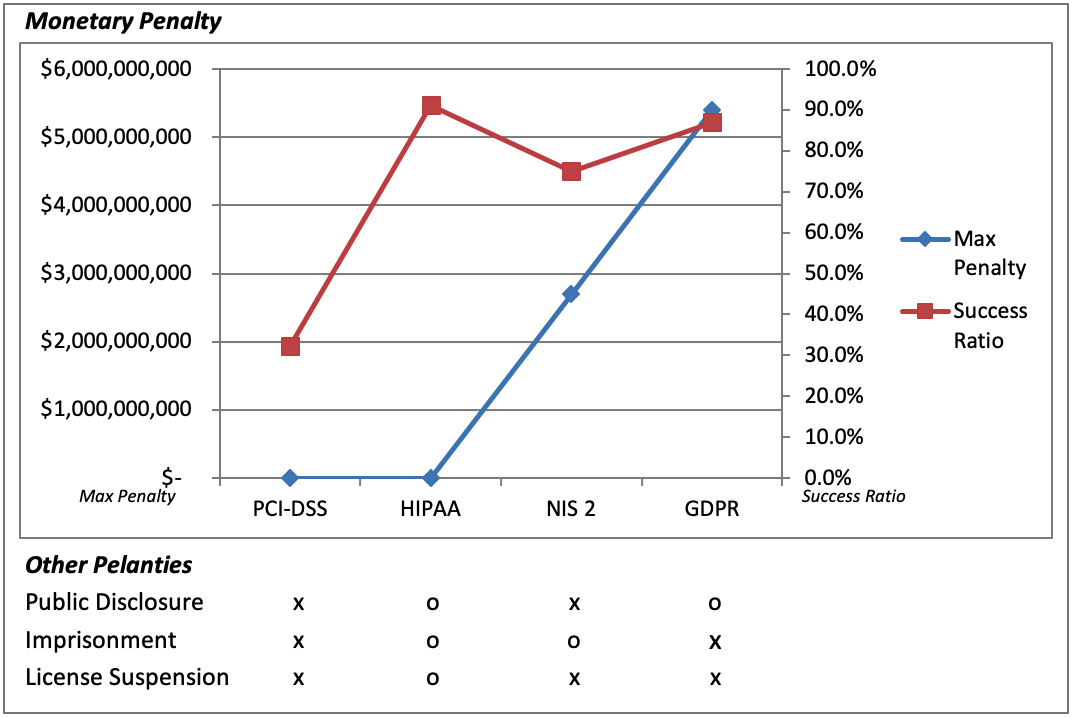}  
\vspace{-0.5em}
        \caption{Comparison of Enforcement Methods, Monetary Penalties, and Success Rates by Framework}
        \label{fig:success rates comparison}
    \end{figure}

What is the relationship between enforcement methods and regulatory success? Table \ref{table:Comparison of Enforcement Methods, Financial Impact, and Success Rates} compares enforcement methods, maximum potential penalties (as business impact), and success rates (from Section II) for each framework, while Fig. \ref{fig:success rates comparison} visualizes the correlation between monetary penalties, enforcement methods, and success rates. 
As illustrated, there appears to be a positive correlation between the severity of financial penalties and the success rates of enforcement. Frameworks with higher monetary penalties (such as GDPR and NIS2) generally demonstrate higher success rates, while those with lower penalties (e.g., PCI DSS) tend to show lower enforcement effectiveness.

    It is worth noting, however, that HIPAA, despite its relatively low financial penalties, employs additional non-monetary enforcement mechanisms such as public disclosure, imprisonment, and license suspension. These measures may enhance its overall enforcement impact beyond what is reflected in financial terms alone.
    
\vspace{-0.5em}
\begin{table}[ht]
\centering

\caption{Comparison of Enforcement Methods, Financial Impact, and Success Rates}
\label{table:Comparison of Enforcement Methods, Financial Impact, and Success Rates}
 
\vspace{-0.5em}
    \begin{tabular}{ m{2em}  m{3cm}  m{1.5cm}  m{1cm}  m{0.6cm} }

    \hline
    &    
    \makecell[c]{Enforcement\\ Methods} &
    \makecell[c]{Potential\\ Max\\ Penalty} & 
    \makecell[c]{Potential\\ Business\\ Impact} &
    \makecell[c]{Success\\ Rates} \\
    \\

    \hline
    PCI DSS&
        \makecell[c]{
        Max penalty \$1.2M/year\\        
        (\$0.1M/month \cite{soroush_pci_2025})\\ 
    }&
    \makecell[c]{\textbf{\$1.2M}\\ {\scriptsize(Max penalty)}}&
    \makecell[c]{0.00089\%}&
    \makecell[c]{\textbf{32\%}}\\

    \hline
    HIPAA &
    \makecell[l]{
       Max penalty \$1.5M/year\\
       per violation type \cite{usdepartment_civil_2013},\\
       public disclosure \cite{usdepartment_summary_2025},\\
       imprisonment \cite{usdepartment_summary_2025},\\
       license suspension \cite{federation_about_2025}
    }&
    \makecell[c]{\textbf{\$8.4M}\\{\scriptsize(4 categories}\\ \raisebox{0.2ex}{$\scriptstyle \times$} {\scriptsize\$2.1M)}}&
    \makecell[c]{0.0062\%}&
    \makecell[c]{\textbf{92\%}}\\

    \hline
    NIS2&
    \makecell[c]{
        Max penalty $\ge$€10M or \\
        2\% global turnover \cite{europeanunion_directive_2022},\\
        imprisonment varies\\ \cite{europeanunion_directive_2022, belgium_law_2019, belgium_law_en_2019}
    }&
    \makecell[c]{\textbf{\$2.7B}\\{\scriptsize(2\% of 134.9B)}}&
    \makecell[c]{2\%}&
    \makecell[c]{\textbf{70\%}}\\
    
    \hline
    GDPR &
    \makecell[c]{
        Max penalty €20M or 4\% \\
        of global turnover,\\ 
        public disclosure \cite{europeanpaliament_regulation_2016}\\
    }&
    \makecell[c]{\textbf{\$5.4B}\\{\scriptsize(4\% of 134.9B)}}&
    \makecell[c]{4\%}&
    \makecell[c]{\textbf{87\%}}\\

    \hline
    \end{tabular}
\end{table}

\vspace{-1.0em}

\subsection{Recommendations for Enhancing PCI DSS}
Drawing on the findings from our analysis, we propose the following recommendations to enhance compliance with PCI DSS.

\subsubsection{Recommendation 1 — Focus on non-monetary penalties rather than higher monetary fines}
PCI DSS is not a legal regulation but a standard established by an industry working group. Because it applies only to organizations that process credit card payments, many organizations have a hard time complying with it given the already high cost. Organizations may decide not to use credit cards if the penalties are too high. Therefore, indirect market pressures, such as public disclosure, are preferable rather than high monetary penalties to encourage organizations to comply.

\subsubsection{Recommendation 2 — Establish an independent monitoring authority} 

The local authority monitoring PCI DSS violations is typically the organization's acquiring bank. Because this authority (the acquiring bank) is commercially tied to the organization, imposing severe penalties on the organization can be difficult for fear of losing clients. Therefore, an independent authority for monitoring violations could strengthen impartiality and help ensure proper PCI DSS compliance.

\section{Conclusion}
\label{conclusion}

PCI DSS is a well-designed standard that contributes to protecting credit card information. However, many organizations do not comply with PCI DSS due to a lack of understanding and weak enforcement. Even when an organization and its acquiring bank are aware that the organization violated PCI DSS standards, it is feasible for them to continue accepting credit cards with only small monetary penalties.

Thus it is imperative to improve authorities’ power to encourage PCI DSS. Authorities should have stronger enforcement powers and be able to provide clear guidelines for ambiguous cases. The structure of central and local authorities should be improved to apply international standards with independent monitoring power instead of using contracted acquiring banks as local authorities. It is also recommended to focus on penalties other than severe monetary ones in order to encourage PCI DSS compliance, rather than measures that risk disrupting card acceptance.

Increasing market pressure is one appropriate approach to PCI DSS enforcement. This includes disclosing PCI DSS compliance status to inform cardholder decisions, thereby potentially affecting the revenues of organizations. An independent authority in charge of this public disclosure could also monitor violations, decide public disclosure cases, respond to unclear cases, and escalate into a central authority.

A central authority could manage, educate, and monitor local authorities. It could also handle escalated cases from local authorities and adjudicate disputes between local authorities and organizations. Empowering central and local authorities appropriately will improve compliance with PCI DSS.

\printbibliography

\appendices

\section{Data Used in Success Rate Calculations}
\begin{table}[ht]
\centering
\caption{HIPAA Enforcement Case Counts by the HHS Office for Civil Rights (2022-2023) \cite{usdepartment_resolution_nd}}
\label{tab:hipaa_cases}
 
\vspace{-0.5em}
    \footnotesize
    \begin{tabular}{ m{1.5em} m{1.5em}  m{7cm}}
    
    \hline
    Year & Total & Case Breakdown  \\
    \hline
    2022 & 22 &
    3/28/2022 - 4 case; 7/14/2022 - 1 case; 7/15/2022 - 11 cases; 8/23/2022 - 1 case; 9/20/2022 - 3 cases; 12/14/2022 - 1 case; 12/15/2022 - 1 case \\

    \hline
    2023 & 13 &
    1/3/2023 - 1 case; 2/2/2023 - 1 case; 5/8/2023 - 1 case; 5/16/2023 - 1 case; 6/5/2023 - 1 case; 6/15/2023 - 1 case; 6/28/2023 - 1 case; 8/24/2023 - 1 case; 9/11/2023 - 1 case; 10/30/2023 - 1 case; 10/31/2023 - 1 case; 11/15/2023 - 1 case; 11/20/2023 - 1 case \\
    \hline
    \end{tabular}
\end{table}

\vspace{-1.5em}
\begin{table}[ht]
  \caption{Estimated 
  Organizations Subject to HIPAA}
  \label{tab:hipaa-scope-note}
\vspace{-0.5em}
  \noindent\begin{minipage}{\columnwidth}
\footnotesize
Between 2022 and 2024, HIPAA applied to over 6,000 hospitals (6,093 in 2023) \cite{american_fast_2025}, 200,000–300,000 physician practices based on a growing physician workforce \cite{association_usphysician_2023,association_usphysician_2024}, and more than 1,200 health insurers. Each hospital typically works with 250–750+ business associates \cite{luoma_business_2025}. Accounting for overlap, an estimated 2.0 to 3.0 million organizations—including about 460,000 covered entities and 1.5–2.5 million unique business associates—were consistently subject to HIPAA compliance during this period.
  \end{minipage}
\end{table}

\vspace{-1.5em}
\begin{table}[h]
\centering

\caption{GDPR Penalty Cases \cite{enforcement_statistics_2023}}
\label{tab:gdpr_penalty} 
\vspace{-0.5em}

    \footnotesize
    \begin{tabular}{ m{1.5em} m{1.5em}  m{6.5cm}}

    \hline
    Year & Total & Monthly Breakdown  \\
    \hline
    2021 & 463 &
    Jan-Dec = 32, 35, 48, 34, 36, 34, 47, 24, 44, 33, 39, 57 \\

    \hline
    2022 & 493 &
    Jan-Dec = 36, 40, 25, 51, 51, 37, 47, 41, 43, 49, 48, 68 \\
    \hline


    \end{tabular}
\end{table}

\vspace{-1.5em}
\begin{table}[ht]
\centering
\caption{Total number of companies in EU states \cite{hithorizons_breakdown_2023}}
\label{tab:eu_companies} 
\vspace{-0.5em}
    \footnotesize
    \begin{tabular}{ m{1em} m{3em}  m{6.5cm}}

    \hline
    Year & Total & Country Breakdown  \\
    \hline
    2021 & 35,430,241 &
    Austria - 589615; Belgium - 843146; Bulgaria - 382086; Croatia - 217553; Cyprus - 83062; Czechia - 1254110; Denmark - 291188; Estonia - 136867; Finland - 429231; France - 4827641; Germany - 3145703; Greece - 880349; Hungary - 949587; Iceland - 44278; Ireland - 368864; Italy - 4462146; Latvia - 138359; Lithuania - 300186; Luxembourg - 41841; Malta - 52095; Netherlands - 2052683; North Macedonia - 65735; Poland - 2665170; Portugal - 1231887; Romania - 934417; Serbia - 213099; Slovakia - 595131; Slovenia - 185655; Spain - 3416248; Sweden - 1031880; Türkiye - 3600429 \\

    \hline
    2022 & 36,715,077 &
    Austria - 583947; Belgium - 888925; Bulgaria - 394135; Croatia - 227408; Cyprus - 87707; Czechia - 1292436; Denmark - 380208; Estonia - 153907; Finland - 442264; France - 5202687; Germany - 3164855; Greece - 917441; Hungary - 976964; Iceland - 44981; Ireland - 389654; Italy - 4579525; Latvia - 145441; Lithuania - 329361; Luxembourg - 45021; Malta - 51506; Netherlands - 2204281; North Macedonia - 66577; Poland - 2675865; Portugal - 1329175; Romania - 974968; Serbia - 205316; Slovakia - 635781; Slovenia - 194876; Spain - 3487503; Sweden - 835543; Türkiye - 3806819 \\
    \hline
    \end{tabular}
\end{table}

\end{document}